\documentclass[11pt]{article}

\usepackage[final]{acl}
\usepackage{fontawesome5}
\usepackage{times}
\usepackage{latexsym}
\usepackage{subcaption}
\usepackage{multirow}
\usepackage[T1]{fontenc}

\usepackage[utf8]{inputenc}

\usepackage{microtype}

\usepackage{inconsolata}
\usepackage{tabularx}
\usepackage{graphicx}
\usepackage{booktabs}
\usepackage{colortbl}
\usepackage{textcomp}
\usepackage{amssymb}
\usepackage{amsmath}
\usepackage{algorithm}
\usepackage{algpseudocode}
\algrenewcommand\algorithmicrequire{\textbf{Input:}}
\algrenewcommand\algorithmicensure{\textbf{Output:}}
\title{\textsc{LexPath}: A Domain-Oriented Multi-Path Framework for Legal Article Retrieval}

\author{
Weixuan Liu \quad
Qingfeng Zhuge\thanks{Corresponding author.} \quad
Xuyang Chen \\
East China Normal University \\
\texttt{qfzhuge@cs.ecnu.edu.cn}
}

\begin{document}
\maketitle
\begin{abstract}
Legal article retrieval is critical for building traceable and reliable legal AI systems, where conclusions must be grounded in specific legal articles.
However, general-purpose retrieval methods rely heavily on lexical or semantic similarity, making it difficult to distinguish legally relevant articles from textually similar but legally inapplicable ones, particularly when they differ in their underlying legal intent.
To bridge this gap, we propose \textsc{LexPath}, a domain-oriented multi-path framework comprising a multi-path retrieval module and an intent-aware reranking module.
The retrieval module combines two complementary domain-specific paths to collect candidate articles: an IRAC-guided sparse path that expands queries with legally informative keywords, and a structure-guided dense path trained with hard negatives derived from legal hierarchy and citation relations.
The reranking module further refines candidate rankings by incorporating the intent compatibility between queries and legal articles.
We evaluate \textsc{LexPath} on three Chinese benchmarks spanning diverse query scenarios and further assess its cross-jurisdictional applicability on a Japanese benchmark.
Experimental results demonstrate that \textsc{LexPath} consistently outperforms lexical, dense, hybrid, and adaptive retrieval-augmented generation (RAG) baselines across all four benchmarks.
Ablation studies further verify the effectiveness of each component. The data and code are available \href{https://github.com/wxliuabigail/LexPath}{here~\faGithub}.

\end{abstract}

\section{Introduction}

\begin{figure}[!t]
    \centering
    \includegraphics[trim=120 90 490 10, clip, width=0.95\linewidth]{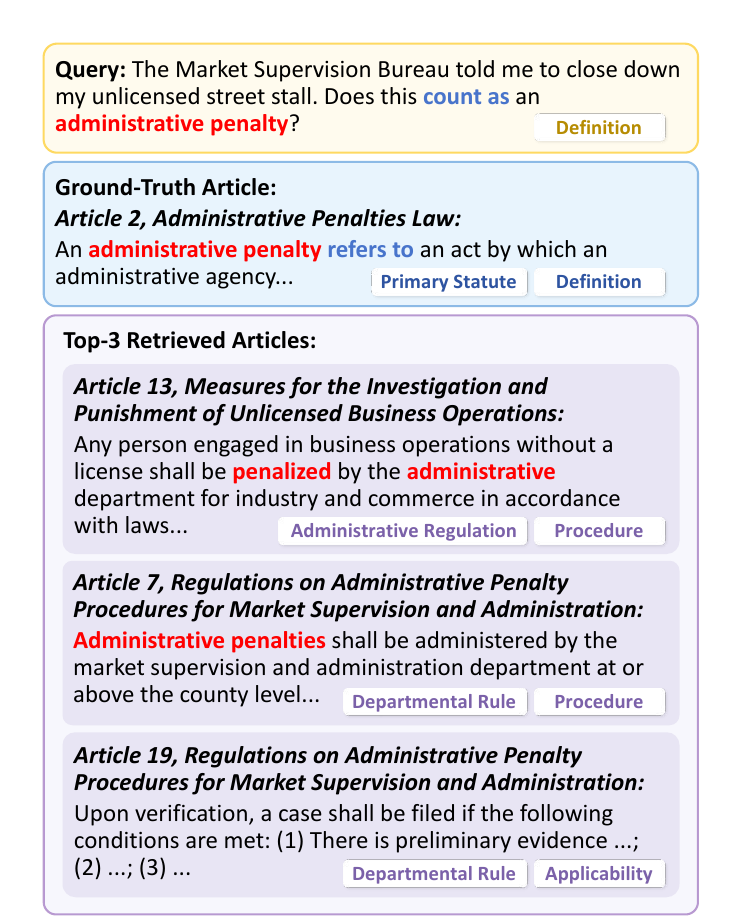}
    \caption{An example where top-ranked articles are textually related but legally inappropriate. Red text denotes keyword matches, and each article is annotated with its legal hierarchy level and intent label.}
    \label{fig:intro}
\end{figure}

Legal article retrieval is crucial for building traceable and reliable legal AI systems~\cite{pipitone2024legalbench}, where conclusions must be supported by specific legal articles.
This task poses significant challenges, even to existing state-of-the-art (SOTA) open-domain solutions, because legal relevance is not equivalent to lexical or semantic similarity~\cite {li2025delta}.
Legal articles that appear similar in wording or topic may involve different legal concepts, authorities, scopes, procedural functions, or conditions of applicability.

Legal article retrieval involves several domain-specific challenges.
First, user queries often describe concrete situations, or use colloquial expressions, while legal articles are written in abstract and terminology-intensive language~\cite{su2024stard}.
Second, legal articles are organized within a legal hierarchy. Even when articles are semantically similar or linked by citations, they may differ in legal authority and may not be equally applicable to a given query~\cite{legislation_law_prc_2023,su2024stard}.
Third, open-domain retrievers may return top-ranked articles that are topically related to the query but do not match its underlying legal intent~\cite{shao2023intent}.

An example in Figure~\ref{fig:intro} shows that legal article retrieval requires more than mere lexical or semantic relevance.
The ground-truth article defines an administrative penalty in a primary statute.
Although the retrieved articles mention administrative penalties, they concern penalty procedures or specific application conditions, and originate from regulations or departmental rules with lower authority than the ground-truth article.

Existing open-domain solutions are far from perfect for this task.
Lexical matching methods~\cite{robertson2009probabilistic,ponte2017language} rely on lexical matching and struggle when users describe concrete scenarios or use colloquial expressions.
Dense embedding methods~\cite{text2vec,bge_embedding,li2023towards,wang2024multilingual,bge-m3} excel at modeling semantic similarity, but struggle to distinguish legal applicability.
Although hybrid methods combine sparse and dense features, they still fail to explicitly incorporate legal-specific features and remain constrained by surface-level textual similarity.
Adaptive retrieval-augmented generation (RAG) methods~\cite{trivedi2023interleaving,yan2024corrective,du2026rag} improve retrieval by reformulating queries or iteratively refining retrieved evidence, but they similarly fail to explicitly model legal-specific relevance.
As a result, existing methods often rank inapplicable but textually similar articles above the truly applicable ones.

To address these challenges, we propose \textsc{LexPath}, a domain-oriented framework consisting of a multi-path retrieval module and an intent-aware reranking module.
The retrieval module first collects candidate articles through two complementary paths: an IRAC-guided sparse path and a structure-guided dense path.
In the sparse path, we propose IRAC-Exp to expand queries with retrieval-oriented keywords.
In the dense path, we propose Struct-Neg to train the embeddings with hierarchy- and citation-aware hard negatives.
Then the candidate ranking is refined by the reranking module that incorporates an intent consistency score.

We evaluate \textsc{LexPath} on three Chinese benchmarks spanning
diverse query scenarios and further assess its cross-jurisdictional
applicability on COLIEE, a Japanese legal retrieval benchmark.
Experiments show that \textsc{LexPath} consistently outperforms lexical, dense, hybrid, and adaptive RAG baselines.
Further ablation studies demonstrate the effectiveness of each component, and efficiency analysis illustrates the trade-off between retrieval quality and latency.

In summary, our contributions are as follows:

\begin{itemize}
    \item We propose \textsc{LexPath}, a domain-oriented multi-path framework for legal article retrieval that jointly models lexical, structural, and intent-level relevance between user queries and legal articles.

    \item We construct StatuteRAG, a legal article retrieval benchmark based on professional training materials, which complements existing benchmarks by covering a distinct application scenario.

    \item Experiments on four benchmarks show that \textsc{LexPath} consistently outperforms lexical, dense, hybrid, and adaptive RAG baselines, and ablation studies further verify the effectiveness of each component.
\end{itemize}

\begin{figure*}
    \centering
    \includegraphics[width=0.92\linewidth,trim = 90 125 310 50]{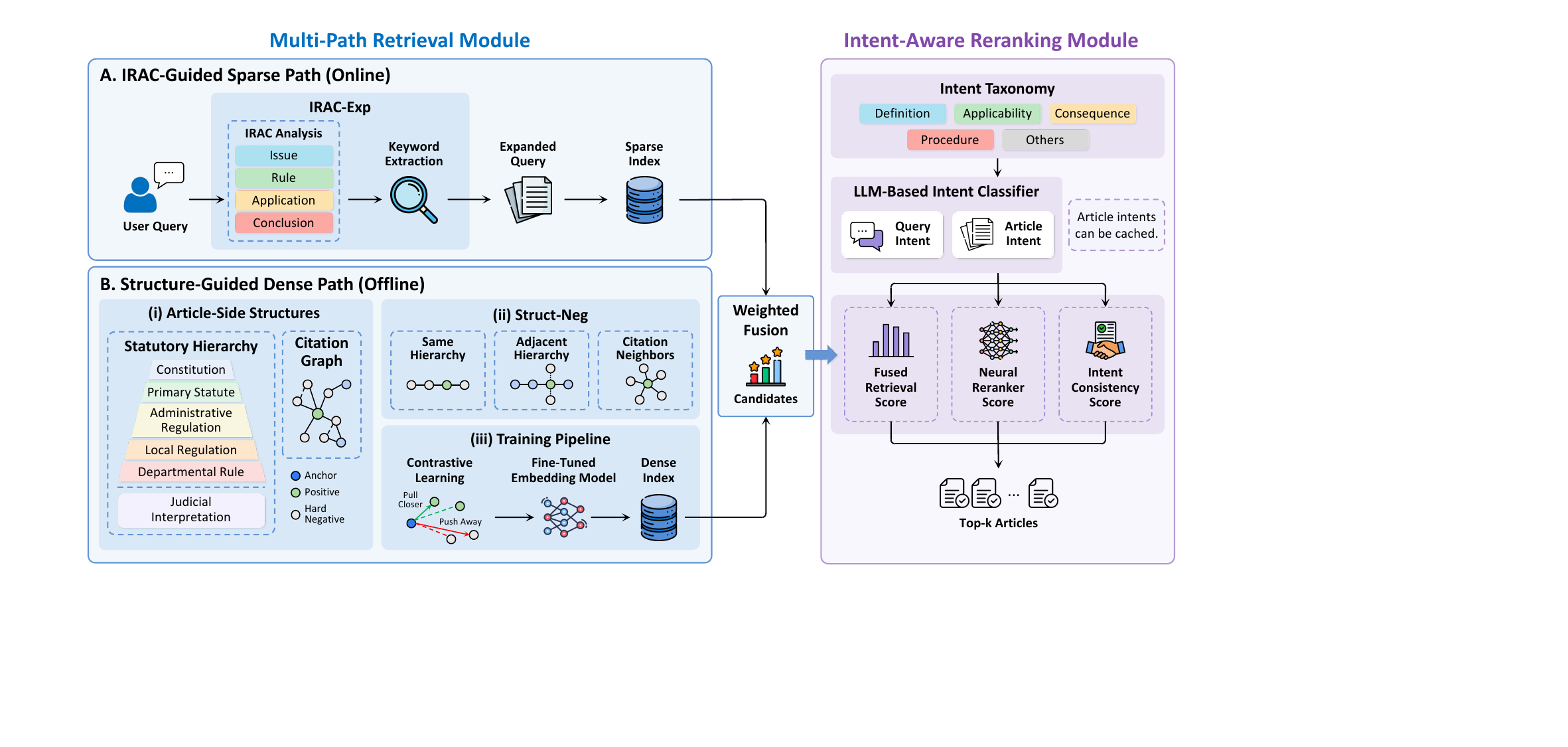}
    \caption{An overview of our framework, \textsc{LexPath}, for legal article retrieval.}
    \label{fig:framework}
\end{figure*}


\section{Related Work}


There is a growing interest in building reliable legal AI systems across different languages~\cite{bhattacharya2019fire,louis2022statutory,goebel2023summary,pipitone2024legalbench,hou2025clerc}.
Legal applications often require conclusions to be traceable to specific legal articles~\cite{pipitone2024legalbench}, making legal article retrieval a fundamental capability for downstream tasks such as legal QA~\cite{fei2024lawbench,louis2024interpretable, li2025legalagentbench}, judgment prediction~\cite{xiao2018cail2018}, and judgment document generation~\cite{su2025judge}.
Despite its importance, legal article retrieval remains less explored than legal case retrieval, which has been extensively studied to identify similarities between judicial precedents~\cite{SAILER,shao2023intent,deng2024element,kim2025legalsearchlm} using domain-specific features.
Recent work has explored incorporating legal structure into statutory article retrieval, such as modeling relations between legal articles with graph-based methods~\cite{louis2023finding}.

Open-domain retrieval methods, including lexical matching~\cite{robertson2009probabilistic,ponte2017language}, dense embedding~\cite{bge_embedding,li2023towards,wang2024multilingual}, and adaptive RAG~\cite{trivedi2023interleaving,yan2024corrective,du2026rag}, are not fully suited to legal article retrieval.
These methods usually do not explicitly model domain-specific factors such as legal terminology, legal hierarchy, and query-article intent consistency, and therefore capture only part of the relevance between user queries and legal articles~\cite{deng2024element,li2025lexrag}.
These limitations motivate our work, which focuses on fine-grained domain-specific relevance modeling for legal article retrieval.

\section{Methodology}

\subsection{Overview}

To address the challenges discussed above, we propose \textsc{LexPath}, a legal-specific multi-path retrieval framework that models lexical, legal structural, and intent-level relevance in a unified manner.

We define legal article retrieval as a task over an article corpus $C$ and a query set $Q$. For each query $q \in Q$, we annotate a ground-truth set of articles $A_q \subseteq C$.
The goal is to develop a retrieval model that, given $q$, ranks the articles in $C$ so that articles in $A_q$ are assigned higher ranks and can be retrieved in the top-$K$ results.
As shown in Figure~\ref{fig:framework}, \textsc{LexPath} first collects candidate articles $R_q \subseteq C$ with $|R_q| > K$ by weighting the scores from two complementary retrieval paths.
Then, the intent-aware reranker refines the candidate ranking in $R_q$ by incorporating the intent consistency score, yielding the final top-$K$ ranked articles.

\subsection{IRAC-Guided Sparse Path}

This component aims to improve sparse retrieval in legal article retrieval.
Lexical matching is important in this task because small wording differences may correspond to different legal concepts, applicable conditions, or legal consequences.
However, the original query may not contain all the keywords needed to retrieve the relevant article.
While a query may describe concrete scenarios or use colloquial expressions, the target article is typically characterized by abstract legal concepts and dense terminology.
To address this issue, we develop \textbf{IRAC-Exp}, which uses IRAC analysis to expand the query with legal keywords for sparse retrieval.


IRAC analysis is a commonly used legal reasoning framework that decomposes a legal problem into Issue, Rule, Application, and Conclusion.
Prior studies have shown that IRAC analysis can improve the performance of LLMs in complex legal reasoning tasks \cite{yu2022legal,servantez2024chain}.
The generated IRAC analysis is not used as the final answer. Instead, it serves as an intermediate representation to expose legal concepts and retrieval-oriented terminology from the original query.

In IRAC-Exp, we first prompt the LLM to perform an IRAC analysis of the user query. We then extract retrieval-oriented legal keywords from the resulting IRAC analysis and append them to the original query for BM25~\cite{robertson2009probabilistic} retrieval.
For the specific prompts used for IRAC analysis and keyword extraction, please refer to Figure~\ref{fig:prompt-irac} and Figure~\ref{fig:prompt-keyword} in Appendix~\ref{sec:prompt}, respectively.


\subsection{Structure-Guided Dense Path}

This component is designed to address the limitation of open-domain embedding models, which often fail to distinguish between semantically similar articles that differ in legal authority, applicability, or scope in a structured legal system.
Specifically, we train the embedding model with contrastive learning and develop \textbf{Struct-Neg}, a structure-aware hard negative mining strategy based on the hierarchical and citation relationships among legal articles.
Given a corpus ${C}$, where each article $a \in {C}$ is associated with a law title $t$, an article number $p$, and article content $x$, we build hierarchy buckets and a citation graph for hard negative mining.

\paragraph{Hierarchy Buckets Construction} To model the hierarchical relationships of the legal system, we assign each article $a$ a discrete hierarchy label $L(a)\in\{0,1,\dots,6\}$ according to its title $t$: $0$ for the Constitution, $1$ for Primary Statute, $2$ for Administrative Regulation, $3$ for Local Regulation or Autonomous Regulations, $4$ for Departmental Rule or Local Government Rule, $5$ for Judicial Interpretations, and $6$ for other normative documents. This taxonomy is constructed with reference to the Legislation Law~\cite{legislation_law_prc_2023}. For retrieval-oriented negative sampling, labels 0–4 are treated as coarse ordered structural tiers, while labels 5 and 6 are non-ordered categories. This ordering is a modeling approximation rather than a complete characterization of legal-effect priority.

Given a query $q$, we partition the retrieved candidate set $R$ into hierarchy buckets according to the hierarchy labels:
\begin{equation}
    B_\ell(q)=\{a\in R \mid L(a)=\ell\}.
\end{equation}

\paragraph{Citation Graph Construction} To capture citation relationships among legal articles, we construct an article-level citation graph based on the corpus. We define a directed graph ${G}=({V},{E})$ with ${V}={C}$ and build an index function $\Phi:(t,p)\mapsto a$ that maps a normalized law title and article number to the corresponding article node.

For each article $a_i$, we extract citation mentions from its content $x$ using pattern matching, including both cross-law citations and internal citations. Each mention is resolved to a target article through title normalization and article-level lookup via $\Phi$, where internal citations are grounded to the title of the source article itself. If a citation mention can be successfully resolved to an article $a_j$, we add a directed edge $(a_i,a_j)\in{E}$. 

\begin{algorithm}
\caption{Workflow of Struct-Neg.}
\label{alg:sampling}
\begin{algorithmic}[1]
\Require Query $q$, positive sample $p^+$, ground-truth article set $A_q$,
article corpus $C$, citation graph $G$, hierarchy mapping $L(\cdot)$,
retrieval depth $K$, negative budget $M$
\Ensure Negative sample set $N_q$

\State $\mathcal{H} \gets \{0,1,2,3,4\}$ \Comment{ordered hierarchy levels}
\State $\ell^+ \gets L(p^+)$
\State $R \gets \mathrm{DenseRetrieve}(q, C, K)$
\State $R \gets R \setminus A_q$
\State Partition $R$ into hierarchy buckets $\{B_\ell\}$ according to $L(\cdot)$

\State $T \gets \{\ell^+\}$

\If{$\ell^+ \in \mathcal{H}$}
    \If{$\ell^+ - 1 \in \mathcal{H}$}
        \State $T \gets T \cup \{\ell^+ - 1\}$
    \EndIf
    \If{$\ell^+ + 1 \in \mathcal{H}$}
        \State $T \gets T \cup \{\ell^+ + 1\}$
    \EndIf
\EndIf

\State $N_q \gets \mathrm{SampleByHierarchy}(\{B_\ell\}, \ell^+, T, M)$

\State $G_q \gets \mathrm{CitationNeighbors}(p^+, G) \setminus A_q$
\State $N_q \gets N_q \cup G_q$

\State \Return $N_q$
\end{algorithmic}
\end{algorithm}

\paragraph{Struct-Neg}
Based on the hierarchy buckets and the citation graph, the workflow of Struct-Neg is shown in Algorithm~\ref{alg:sampling}.
For each query, we first determine the hierarchy label of its positive article and retrieve the top-$K$ most similar articles from the corpus as candidates.
We then sample negatives from the same hierarchy category, adjacent
levels when applicable, and the citation graph to construct multi-source hard negatives.
In contrastive learning, these negatives train the embedding model to better differentiate articles that are semantically similar but possess distinct hierarchical or citation relations.

\subsection{Multi-Path Score Fusion}

To effectively integrate lexical and semantic features, we combine the scores of the sparse and dense paths through weighted fusion.
Let $r_{\mathrm{sp}}(q,a)$ and $r_{\mathrm{de}}(q,a)$ denote the raw sparse and dense relevance scores between query $q$ and article $a$, respectively.
Since the two scores are produced in different value ranges, we first normalize them as
\begin{equation}
\begin{aligned}
\bar{r}_m(q,a)
&=
\frac{1}{2}
+
\frac{1}{\pi}\arctan\!\bigl(r_m(q,a)\bigr), \\
&\qquad m \in \{\mathrm{sp}, \mathrm{de}\}.
\end{aligned}
\end{equation}
where $\bar{r}_{m}(q,a)$ denotes the normalized score of retrieval path $m$.
The final fused score is then computed as
\begin{equation}
r(q,a)=\alpha\,\bar{r}_{\mathrm{sp}}(q,a)+(1-\alpha)\,\bar{r}_{\mathrm{de}}(q,a),
\end{equation}
where $\alpha\in[0,1]$ controls the relative contribution of the sparse and dense paths.
The fused scores rank the articles, and a pool of highly ranked articles $R_q$ is used as candidates for reranking.

\subsection{Intent-Aware Reranker}
Although the multi-path retrieval module retrieves high-quality candidate articles, it does not explicitly model intent consistency between user queries and legal articles.
In real-world applications, such intent priors are usually unavailable due to the scarcity of domain annotations.
To address this, we leverage few-shot prompting to identify intent labels and refine the final rankings.

\paragraph{Intent Taxonomy} 
We propose a task-oriented intent taxonomy for legal article retrieval, drawing on previous research on legal case retrieval and the typology of legal rules~\cite{nazarenko2018annotation,shao2023intent}: (1) \textbf{Definition}, which seeks the interpretation or clarification of legal concepts and terms;
(2) \textbf{Applicability}, which concerns whether and under what conditions the legal article applies to the facts of the case;
(3) \textbf{Consequence}, which concerns legal liabilities, sanctions, or other legal effects triggered by a factual situation or violation;
(4) \textbf{Procedure}, which concerns procedural requirements, execution steps, and related processes;
and (5) \textbf{Others}, a residual category covering articles that do not fit the above intents, such as legislative purpose clauses.

\paragraph{Intent-Aware Reranking} 
We use few-shot prompting with the LLM to assign both queries and legal articles to one of the predefined intent labels.
Separate few-shot prompt templates are designed for queries and legal articles, as described in Figure~\ref{fig:prompt-query} and Figure~\ref{fig:prompt-article}, respectively.
Given a query $q$ and its candidate article set $R_q$, we use an LLM-based intent classifier to assign intent labels to the query and each candidate article $a \in R_q$:
\begin{equation}
\begin{aligned}
    \hat{y}_q &= \mathrm{LLM}_{\text{intent}}(q), \\
    \hat{y}_a &= \mathrm{LLM}_{\text{intent}}(a), \quad \forall a \in R_q,
\end{aligned}
\end{equation}
where $\hat{y}_q$ and $\hat{y}_a$ are the predicted intent labels.
To verify intent reliability for reranking, we manually evaluated 100 queries and 100 articles. The LLM classifier achieved 81\% and 86\% agreement with human labels, with Cohen's $\kappa$ scores of 0.77 and 0.82 for queries and articles, respectively.

Given a query $q$ and its candidate article set $R_q=\{a_1,\dots,a_W\}$, we define the intent consistency score for each candidate article $a \in R_q$ as
\begin{equation}
s_i(q,a)=
\begin{cases}
1, & \hat{y}_q=\hat{y}_a,\\
0, & \text{otherwise},
\end{cases}
\end{equation}
where $\hat{y}_q$ and $\hat{y}_a$ are the predicted intent labels of the query and article. 

The final reranking score is then computed by
\begin{equation}
s(q,a)=\lambda_1 s_r(q,a)+\lambda_2 s_p(q,a)+\lambda_3s_i(q,a),
\end{equation}
where $s_r(q,a)$ denotes the neural reranker score, and $s_p(q,a)=r(q,a)$ denotes the fused retrieval score in
Equation~(3).
\begin{table*}[t]
\centering
\small
\caption{Main retrieval results on STARD, LexRAG, StatuteRAG, and COLIEE. ``R'' stands for Recall, and ``N'' stands for NDCG. Best and second-best results are shown in \textbf{bold} and \underline{underlined}, respectively.}
\begin{tabularx}{\linewidth}{
l
!{\hspace{1.2pt}\vrule width 0.5pt\hspace{1.2pt}}
>{\centering\arraybackslash}X
>{\centering\arraybackslash}X
!{\hspace{1.2pt}\vrule width 0.5pt\hspace{1.2pt}}
>{\centering\arraybackslash}X
>{\centering\arraybackslash}X
!{\hspace{1.2pt}\vrule width 0.5pt\hspace{1.2pt}}
>{\centering\arraybackslash}X
>{\centering\arraybackslash}X
!{\hspace{1.2pt}\vrule width 0.5pt\hspace{1.2pt}}
>{\centering\arraybackslash}X
>{\centering\arraybackslash}X
}
\toprule
\textbf{Dataset}
& \multicolumn{2}{c}{\textbf{STARD}}
& \multicolumn{2}{c}{\textbf{LexRAG}}
& \multicolumn{2}{c}{\textbf{StatuteRAG}}
& \multicolumn{2}{c}{\textbf{COLIEE}} \\
\cmidrule(l{4pt}r{8pt}){2-3}
\cmidrule(l{4pt}r{8pt}){4-5}
\cmidrule(l{4pt}r{8pt}){6-7}
\cmidrule(l{4pt}r{8pt}){8-9}

\textbf{Model}
& \textbf{R@5} & \textbf{N@5}
& \textbf{R@5} & \textbf{N@5}
& \textbf{R@5} & \textbf{N@5}
& \textbf{R@5} & \textbf{N@5} \\

BM25
& 37.54 & 33.63
& 13.69 & 9.73
& 72.34 & 62.95
& 73.85 & 66.26 \\

QL
& 36.49 & 31.33
& 12.54 & 8.85
& 71.25 & 62.17
& 71.10 & 61.73 \\

\midrule
text2vec-base-chinese
& 36.41 & 31.39
& 12.52 & 9.20
& 57.33 & 48.29
& 39.91 & 31.68 \\

bge-large-zh-v1.5
& 45.95 & 39.80
& 20.73 & 15.90
& 69.78 & 57.70
& 47.71 & 39.27 \\

gte-Qwen2-1.5B
& 32.49 & 29.00
& 11.51 & 8.44
& 71.06 & 56.22
& 46.33 & 37.16 \\

bge-m3
& 47.47 & 41.08
& 23.60 & 17.61
& 72.53 & 59.53
& 73.39 & 65.10 \\

multilingual-e5-large
& 44.72 & 40.01
& 17.95 & 13.09
& 73.26 & 61.22
& 72.02 & 64.08 \\

\midrule
ChatLaw\_Text2Vec
& 14.70 & 11.23
& 7.25 & 4.79
& 15.02 & 10.41
& 12.39 & 6.96 \\

SAILER\_zh
& 26.58 & 22.76
& 7.05 & 5.02
& 51.47 & 42.39
& 23.85 & 21.16 \\

\midrule

Graph-enhanced bge-large-zh-v1.5
& 47.42 & 40.86
& 22.71 & 17.33
& 69.78 & 57.82
& 53.67 & 46.73 \\

LLeQA-style bge-large-zh-v1.5
& 36.65 & 31.60
& 13.50 & 9.89
& \underline{78.21} & \underline{68.45}
& 70.64 & 62.46 \\

\midrule
BM25 + bge-large-zh-v1.5
& 43.98 & 39.83
& 19.64 & 14.54
& 75.82 & 62.60
& 70.18 & 62.40 \\

BM25 + bge-m3
& 45.68 & 40.65
& 20.51 & 15.83
& 77.84 & 65.45
& 79.82 & 71.82 \\

\midrule
BM25 + jina-reranker
& 44.15 & 38.08
& 16.33 & 10.90
& 75.27 & 64.76
& 76.15 & 67.29 \\

bge-m3 + jina-reranker
& \underline{48.88} & 42.35
& \underline{23.96} & \underline{17.87}
& 73.26 & 60.69
& 78.44 & 67.17 \\

BM25 + bge-m3 + jina-reranker
& 48.77 & \underline{42.97}
& 23.22 & 17.59
& 73.44 & 62.50
& \underline{81.19} & \underline{72.50} \\

\midrule
IRCoT
& 30.98 & 27.16
& 10.49 & 7.24
& 66.67 & 57.66
& 70.64 & 61.56 \\

CRAG
& 39.05 & 32.74
& 12.66 & 9.09
& 71.25 & 61.56
& 73.39 & 63.54 \\

A-RAG
& 30.19 & 28.24
& 15.64 & 13.90
& 45.42 & 42.87
& 75.69 & 66.46 \\

\midrule
\textsc{LexPath}
& \textbf{56.31} & \textbf{51.68}
& \textbf{32.45} & \textbf{25.23}
& \textbf{93.04} & \textbf{81.98}
& \textbf{86.24} & \textbf{77.62} \\

\bottomrule
\end{tabularx}

\label{tab:main-results}
\end{table*}


\section{Experiments}
\subsection{Setup}
\paragraph{Datasets}

We evaluate \textsc{LexPath} on four legal article retrieval benchmarks covering diverse query sources and legal scenarios.
The three Chinese
benchmarks cover scenarios based on real-world public legal consultations, multi-turn public legal consultation dialogues, and professional training
materials.
We further evaluate cross-jurisdictional applicability on a Japanese benchmark based on bar examination questions.

\begin{itemize}
    \item \textbf{STARD}~\cite{su2024stard} contains 1,543 real-world legal consultation queries and 55,348 candidate legal articles, with reference article annotations but no reference answers.
    \item \textbf{LexRAG}~\cite{li2025lexrag} contains 1,013 multi-turn public legal consultation dialogues and 17,228 candidate legal articles, with reference articles, reference answers, and evaluation keywords for each query.
    \item \textbf{StatuteRAG} is our self-constructed benchmark adapted from professional training materials in market supervision and administration. It contains 1,361 queries and 56,982 candidate legal articles, with reference article annotations and reference answers for each query. Construction details are provided in Appendix~\ref{sec:statuterag}.
    \item \textbf{COLIEE 2026 Task 3}~\citep{coliee2026task3} is based on Japanese bar examination questions. The released training set contains 965 queries with relevant article annotations and entailment labels, together with 725 candidate Japanese Civil Code articles. The official test set contains 82 unlabeled queries. We use the statute-retrieval component of the task.
\end{itemize}

\paragraph{Baselines}
We evaluate \textsc{LexPath} against the following baselines:
(1) Lexical matching: BM25~\cite{robertson2009probabilistic}, QL~\cite{ponte2017language};
(2) Open-domain embedding: text2vec-base-chinese~\cite{text2vec}, bge-large-zh-v1.5~\cite{bge_embedding}, gte-Qwen2-1.5B~\cite{li2023towards}, bge-m3~\cite{bge-m3}, multilingual-e5-large~\cite{wang2024multilingual};
(3) Legal-domain embedding: ChatLaw\_Text2Vec~\cite{ChatLaw}, SAILER\_zh~\cite{SAILER};
(4) Legal-specific retrieval: Graph-enhanced bge-large-zh-v1.5~\cite{louis2023finding} and LLeQA-style bge-large-zh-v1.5~\cite{louis2024interpretable};
(5) Hybrid score fusion: BM25 + bge-large-zh-v1.5, BM25 + bge-m3;
(6) Controlled neural reranking: BM25 + jina-reranker-v2-base-multilingual~\cite{jina2025jinarerankerv2}, bge-m3 + jina-reranker-v2-base-multilingual, BM25 + bge-m3 + jina-reranker-v2-base-multilingual;
and (7) Adaptive RAG: IRCoT~\cite{trivedi2023interleaving}, CRAG~\cite{yan2024corrective}, A-RAG~\cite{du2026rag}.
For the legal-specific retrieval baselines, we adapt their retrieval components to our benchmark settings.
For adaptive RAG baselines, we adapt their query reformulation or retrieval-control modules to legal article retrieval while keeping the same base retriever and candidate pool for fairness.

\paragraph{Metrics}
For evaluation, we use standard ranking metrics, including Recall@$K$ and NDCG@$K$, with $K=5$.
We report small-$K$ results because they better reflect the practical application in the legal domain, where the system is expected to precisely reference legal articles, and a larger candidate set may increase noise.

\paragraph{Implementation Details}


Each dataset is split into training, development, and test sets in a 7:1:2 ratio.
For StatuteRAG, we split at the seed-question level to avoid data leakage. For COLIEE, since gold relevant-article annotations for the official test set are not released, we construct the training, development, and test splits from the 965 released training queries using the same 7:1:2 ratio. Retrieval experiments are conducted over the full article corpus, with a candidate pool size of 20.
We use Qwen2.5-7B-Instruct~\cite{qwen2.5} for query expansion, Qwen3-8B~\cite{qwen3technicalreport} for few-shot intent classification, jina-reranker-v2-base-multilingual~\cite{jina2025jinarerankerv2} for neural reranking, and Milvus~\cite{wang2021milvus} for index management.
For the dense path, we fine-tune bge-large-zh-v1.5~\cite{bge_embedding}
with FlagEmbedding for 5 epochs using a batch size of 16 and a learning
rate of $1\times10^{-5}$, and merge the fine-tuned checkpoint with the
original checkpoint at a 5:5 ratio.
For score fusion and reranking, we first tune $\alpha$ and fix it at
$0.4$, and then tune $\lambda_3$ on the development split while fixing
$\lambda_1=\lambda_2=1.0$.
The resulting $\lambda_3$ values are $1.5$, $0.3$, $1.5$, and $1.5$ for
STARD, LexRAG, StatuteRAG, and COLIEE respectively.
For COLIEE, since all candidate articles are drawn from the Japanese
Civil Code and no comparable statutory hierarchy is available, the
dense path reduces to standard contrastive learning with retrieved hard
negatives.
Parameter analysis is provided in Section~\ref{sec:para}.
All experiments are implemented in Python 3.12 with Hugging Face Transformers~\cite{wolf2020transformers}, vLLM~\cite{kwon2023efficient}, and the OpenAI Python library~\cite{openai_python}, and run on a single NVIDIA A6000 GPU with 48GB of memory.
Dense-path training takes less than 0.5 GPU hours.

\subsection{Main Results}

The main results on STARD, LexRAG, StatuteRAG, and COLIEE are presented in Table~\ref{tab:main-results}. We have the following observations:
(1) \textsc{LexPath} consistently outperforms the strongest baseline, improving Recall@5 by 7.43, 8.49, 14.83, and 5.05 on the four benchmarks, respectively, showing the effectiveness of multi-level legal relevance modeling.
(2) Legal-domain embeddings do not always outperform open-domain embeddings, as document-level representations may not necessarily be suitable for article-level retrieval.
(3) Both hybrid retrieval with reranking and adaptive RAG yield limited and inconsistent performance gains. Relying solely on general-purpose strategies such as combining open-domain retrievers, reformulating queries, or expanding to multi-turn retrieval is inherently insufficient without domain-oriented relevance modeling.
(4) All baseline methods underperform on LexRAG, largely due to its multi-turn consultation setting. We further analyze these challenges in Appendix~\ref{sec:chall}. In contrast, the relatively high scores on StatuteRAG partly reflect its mostly single-gold setting and more explicit professional queries. Furthermore, the impact of retrieval quality on downstream legal QA is explored in Appendix~\ref{sec:legalqa}.

\subsection{Ablation Studies}

We conduct ablation studies on three Chinese benchmarks to show the effectiveness of each component of \textsc{LexPath}, as shown in Table~\ref{tab:abla}, including
the following variants: 
(1) w/o Reranker: removing the intent-aware reranker;
(2) w/o Intent Score: removing the intent consistency score in the reranker, leaving the neural reranker score and prior retrieval score used for reranking;
(3) w/o Sparse Path: removing the sparse path, where only the dense path is used for retrieving candidates;
(4) w/o IRAC-Exp: removing the IRAC-Exp in the sparse path, where only BM25 is used for hybrid score fusion;
(5) w/o Dense Path: removing dense path, where only the sparse path is used for retrieving candidates;
(6) w/o Struct-Neg: removing Struct-Neg and using only the original dense retriever for hybrid score fusion.

\begin{table}[h]
\centering
\small
\setlength{\tabcolsep}{5pt}
\caption{Ablation results on three Chinese benchmarks measured by Recall@5.}
\begin{tabular}{l|ccc}
\toprule
\textbf{Method} & \textbf{STARD} & \textbf{LexRAG} & \textbf{StatuteRAG} \\
\midrule
\textsc{LexPath} & \textbf{56.31} & \textbf{32.45} & \textbf{93.04} \\
w/o Reranker & 55.34 & 31.16 & 90.84 \\
w/o Intent Score & 54.69 & \underline{31.76} & \underline{92.67} \\
w/o Sparse Path & \underline{55.66} & 30.97 & \underline{92.67} \\
w/o IRAC-Exp & 55.02 & 29.09 & 91.94 \\
w/o Dense Path & 51.46 & 22.49& 84.98 \\
w/o Struct-Neg & 55.02 & 31.26 & 92.31 \\
\bottomrule
\end{tabular}
\label{tab:abla}
\end{table}

The ablation results show that different components contribute to \textsc{LexPath} at different stages of the retrieval pipeline.
(1) Removing the dense path causes the largest performance drop across three benchmarks, indicating that the dense path serves as the main recall backbone of \textsc{LexPath}.
(2) Removing IRAC-Exp leads to a larger drop than removing the entire sparse path, suggesting that an unexpanded sparse path may introduce noisy lexical matches, whereas IRAC-Exp makes sparse evidence more complementary to the dense path.
(3) Removing Struct-Neg results in a moderate drop, showing that hierarchy- and citation-aware hard negatives help the dense path distinguish semantically similar but legally different articles.
(4) Removing either the reranker or the intent-consistency score leads
to smaller yet consistent drops, suggesting that intent-aware reranking is useful for distinguishing legally applicable articles among candidate articles.

Although the LLM-based components provide consistent overall gains,
they may also introduce errors through noisy query expansion or incorrect
intent prediction.
We provide a detailed analysis of these failure modes
and their effects on reranking in Appendix~\ref{sec:llm-error-analysis}.

\subsection{Design Choice Analysis}

To further examine the key design choices in \textsc{LexPath}, we conduct additional experiments across three Chinese benchmarks, focusing on IRAC-Exp in the sparse path and Struct-Neg in the dense path.

\paragraph{Comparison of Query Expansion Strategies}
We compare IRAC-Exp with representative query reformulation methods, including direct query rewriting (QR)~\cite{ma2023query}, Query2Doc~\cite{wang2023query2doc}, and HyDE~\cite{gao2023precise}. 
As shown in Table~\ref{tab:design}, IRAC-Exp achieves the best performance among all expansion methods.
This suggests that explicitly decomposing a legal query into issue, rule, application, and conclusion helps expose retrieval-oriented legal terminology that is often missing from the original query.
Compared with generic expansion methods, IRAC-Exp provides more task-specific lexical evidence for legal article retrieval.

\begin{table}[h]
\centering
\small
\setlength{\tabcolsep}{5pt}
\caption{Comparison of query expansion and negative mining strategies measured by Recall@5.}
\begin{tabular}{l|ccc}
\toprule

\textbf{Method} & \textbf{STARD} & \textbf{LexRAG} & \textbf{StatuteRAG} \\
\midrule
BM25 & 37.54 & 13.69 & 72.34 \\
\quad + QR & 41.42 & 18.15 & 71.79 \\
\quad + Query2Doc & 44.98 & \underline{21.99} & 76.56 \\
\quad + HyDE & \underline{45.63} & 19.33 & \underline{79.12} \\
\quad + IRAC-Exp & \textbf{47.57} & \textbf{24.65} & \textbf{81.68} \\
\midrule
bge-large-zh-v1.5 & 45.95 & 20.73 & 69.78\\
\quad + Random & 53.40 & \underline{29.68} & \underline{87.55}\\
\quad + ANN & \underline{54.05} & 28.50 & 86.45 \\
\quad + Struct-Neg & \textbf{55.34} & \textbf{30.18} & \textbf{89.74} \\
\bottomrule
\end{tabular}
\label{tab:design}
\end{table}






\paragraph{Comparison of Hard Negative Mining Strategies}
We further compare Struct-Neg with random and approximate nearest neighbor (ANN)-based hard negative mining strategies for dense retriever training. Random negatives provide weak supervision, while ANN-based negatives mainly capture semantic similarity and may overlook legal hierarchy or citation relations. In contrast, Struct-Neg achieves the best performance by sampling negatives from the same or adjacent hierarchy levels and citation-related articles. This indicates that structure-aware negatives provide more informative supervision for distinguishing semantically similar but legally different articles.

\subsection{Parameter Analysis}
\label{sec:para}

\begin{figure}[t]
    \centering
    \captionsetup{skip=2pt}

    \begin{subfigure}[t]{0.49\linewidth}
        \centering
        \includegraphics[trim=10 10 0 10, clip, width=\linewidth]{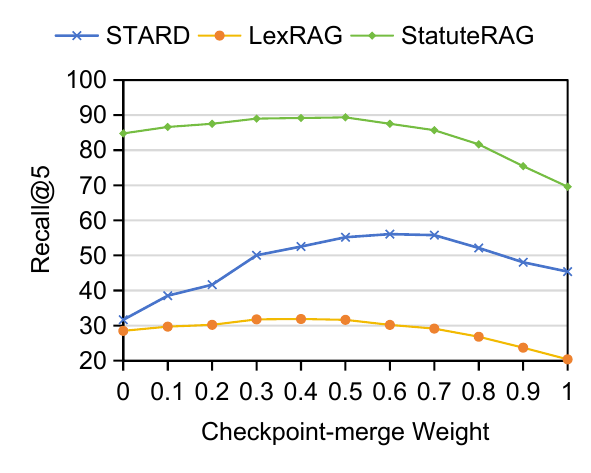}
        \label{fig:cp}
    \end{subfigure}


    \begin{subfigure}[t]{0.49\linewidth}
        \centering
        \includegraphics[trim=10 5 0 10, clip, width=\linewidth]{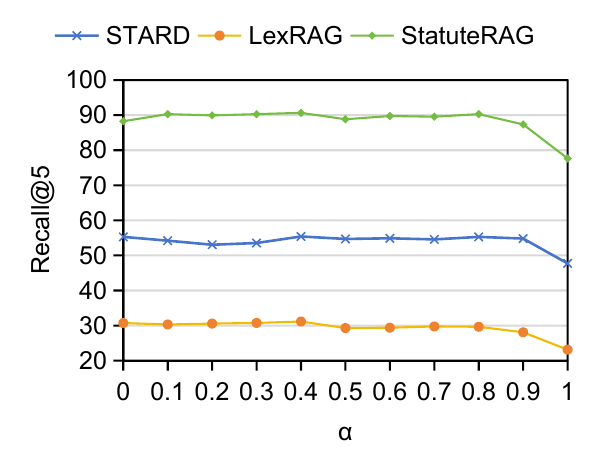}
        \label{fig:alpha}
    \end{subfigure}
    \hfill
    \begin{subfigure}[t]{0.49\linewidth}
        \centering
        \includegraphics[trim=10 5 0 10, clip, width=\linewidth]{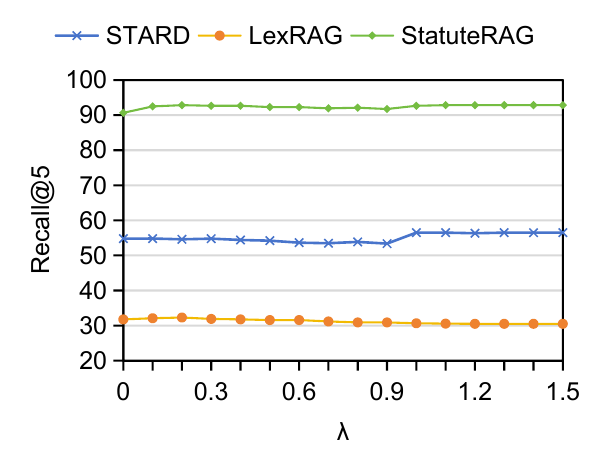}
        \label{fig:lambda}
    \end{subfigure}

    \vspace{-0.8em}
    \caption{Parameter analysis of \textsc{LexPath} by varying the checkpoint-merge weight, sparse-path weight $\alpha$, and intent-consistency score weight $\lambda_3$.}
    \label{fig:para-analysis}
\end{figure}

In this section, we analyze the sensitivity of three key parameters: the checkpoint-merge weight, the sparse-path weight $\alpha$, and the intent-consistency score weight $\lambda_3$. We report Recall@5 in Figure~\ref{fig:para-analysis}.

Across the three benchmarks, moderate checkpoint-merge weights generally perform better than using either checkpoint alone, suggesting that checkpoint merging balances general semantic ability and task-specific structural signals. Performance also remains stable within a moderate range of $\alpha$, confirming the complementarity between sparse and dense paths. When $\alpha$ is close to 1.0, performance drops, indicating that sparse matching alone is insufficient for legal article retrieval. Finally, \textsc{LexPath} is relatively insensitive to $\lambda_3$, as moderate intent-consistency score weights provide useful reranking signals, while overly large weights bring no clear additional gains.
The similar trends across datasets and the broad stable ranges further mitigate concerns about overfitting to the relatively small development sets.
\subsection{Efficiency Analysis}

Table~\ref{tab:efficiency} reports the accuracy--efficiency trade-off on STARD, which we use as a representative dataset because the same retrieval pipeline and number of online LLM calls are used across benchmarks.
The full \textsc{LexPath} achieves the best retrieval performance at a higher inference cost, making it suitable for accuracy-oriented settings.
In contrast, the dense-only configuration requires no online LLM calls and provides a latency-oriented alternative with substantially lower inference cost.
These results show that \textsc{LexPath} can support different deployment requirements by trading off retrieval quality against computational cost.

\begin{table}[h]
\centering
\small
\renewcommand{\arraystretch}{0.95}
\caption{Efficiency analysis on STARD.}
\begin{tabular}{l@{\hspace{4pt}}|@{\hspace{4pt}}ccc}
\toprule

\textbf{Method} & \textbf{R@5} & \textbf{\# LLM Call} & \textbf{Latency (s)} \\
\midrule
BM25 & 37.54 & 0 & 0.33 \\
bge-large-zh-v1.5 & 45.95 & 0 & 0.30 \\
\midrule
\textsc{LexPath} & 56.31 & 3 & 10.12 \\
w/o Reranker & 55.34 & 2 & 6.79 \\
w/o Sparse Path & 55.66 & 1 & 3.61 \\
Sparse Path & 47.57 & 2 & 6.78 \\
Dense Path & 55.34 & 0 & 0.30 \\
\bottomrule
\end{tabular}
\label{tab:efficiency}
\end{table}

\subsection{Case Study}
We present two representative cases to illustrate typical errors made by open-domain retrievers. Please refer to Appendix~\ref{sec:case}.

As shown in Figure~\ref{fig:case1}, the first case illustrates a typical error made by sparse retrievers.
The top-3 articles retrieved by the sparse retriever contain surface keyword matches such as ``Food Safety Law'', but they miss the more informative phrase ``serious circumstances'', which reflects the query's definition-seeking intent.
In addition, sparse retrieval fails to distinguish articles with different hierarchy levels.
In contrast, \textsc{LexPath} ranks the ground-truth article first and assigns higher ranks to articles from more appropriate legal levels.

As shown in Figure~\ref{fig:case2}, the second case illustrates a typical error made by dense retrievers.
The top-3 articles retrieved by the dense retriever are semantically close to the query. Still, they give insufficient attention to the keyword ``suspend'' and fail to distinguish it from the related but legally different term ``terminate''.
They also overlook differences in legal hierarchy.
In contrast, \textsc{LexPath} better captures the keyword ``suspend'', ranks the ground-truth article first, and promotes articles from more appropriate legal levels.

\section{Conclusion}

We proposed \textsc{LexPath}, a domain-oriented framework for legal article retrieval that captures relevance through lexical matching, semantic retrieval, and intent-level consistency.
Specifically, \textsc{LexPath} combines two complementary paths: the sparse path enriched by IRAC-Exp with legal terminology, and the dense path enhanced by Struct-Neg with legal structure-based hard negatives.
The candidate articles are then refined by an intent-aware reranker.
Experiments on four benchmarks show that \textsc{LexPath} consistently outperforms lexical, dense, hybrid, and adaptive RAG baselines, with ablation studies confirming the effectiveness of each component.
Results on COLIEE provide preliminary evidence of cross-jurisdictional applicability, while broader evaluation across additional legal systems remains future work.

\section*{Limitations}

There are several limitations in this work that we
plan to investigate and address in future work:

\begin{itemize}
    \item Although the results on COLIEE provide preliminary evidence of
    cross-jurisdictional applicability, our evaluation is still limited to
    Chinese and Japanese legal systems. Broader evaluation across
    additional jurisdictions and languages is needed to further assess
    the generalizability of \textsc{LexPath}.

    \item The proposed framework uses LLMs for query expansion and intent classification, which inevitably adds inference cost.
    This cost can be partly mitigated because most article-side information can be precomputed and cached.
    Future work may explore more efficient query-side modeling.

    \item This study focuses on retrieval-side optimization and uses a task-oriented intent taxonomy.
    While the taxonomy captures common functional roles of legal articles, some queries and articles may involve multiple or implicit intents.
    Future work may refine the taxonomy and extend the domain-oriented design to legal systems where retrieval, legal reasoning, evidence verification, and answer generation are jointly optimized.
\end{itemize}

\section*{Ethics Statement}
Our proposed dataset is adapted from an existing publicly available professional training manual with appropriate authorization for non-commercial research use.
We properly cite the source and ensure that our use is consistent with both the authorized scope and the original educational purpose of the manual.
The source manual is a published professional training resource that has undergone editorial review and provides well-structured questions, explanations, and legal article references.
The constructed dataset is manually verified by the authors, who are native Chinese speakers and were instructed by legal professionals on the annotation criteria before verification.
No external annotators were involved, and the verification process does not require reviewing private user data, sensitive personal information, or harmful content.
Therefore, our verification focuses on the quality of the conversion process, including query completeness, answer consistency, and article support.
The verification process takes approximately 16 working hours per author.

\section*{Acknowledgments}
This work is partially supported by NSFC 62372183 and STCSM 24LZ1400500.

\bibliography{custom}

@misc{coliee2026task3,
  title        = {{COLIEE 2026 Task 3: Statute Retrieval and Entailment for Yes/No Questions}},
  author       = {{COLIEE}},
  year         = {2026},
  howpublished = {\url{https://coliee.org/COLIEE2026/tasks/task3}},
  note         = {Accessed: 2026-08-24}
}

@inproceedings{louis2024interpretable,
  title={Interpretable long-form legal question answering with retrieval-augmented large language models},
  author={Louis, Antoine and Van Dijck, Gijs and Spanakis, Gerasimos},
  booktitle={Proceedings of the AAAI conference on artificial intelligence},
  volume={38},
  number={20},
  pages={22266--22275},
  year={2024}
}

@inproceedings{louis2023finding,
  title={Finding the law: Enhancing statutory article retrieval via graph neural networks},
  author={Louis, Antoine and Van Dijck, Gijs and Spanakis, Gerasimos},
  booktitle={Proceedings of the 17th Conference of the European Chapter of the Association for Computational Linguistics},
  pages={2761--2776},
  year={2023}
}

@inproceedings{wolf2020transformers,
  title={Transformers: State-of-the-art natural language processing},
  author={Wolf, Thomas and Debut, Lysandre and Sanh, Victor and Chaumond, Julien and Delangue, Clement and Moi, Anthony and Cistac, Pierric and Rault, Tim and Louf, R{\'e}mi and Funtowicz, Morgan and Davison, Joe and Shleifer, Sam and von Platen, Patrick and Ma, Clara and Jernite, Yacine and Plu, Julien and Xu, Canwen and Le Scao, Teven and Gugger, Sylvain and Drame, Mariama and Lhoest, Quentin and Rush, Alexander},
  booktitle={Proceedings of the 2020 conference on empirical methods in natural language processing: system demonstrations},
  pages={38--45},
  year={2020}
}

@inproceedings{wang2021milvus,
  title={Milvus: A purpose-built vector data management system},
  author={Wang, Jianguo and Yi, Xiaomeng and Guo, Rentong and Jin, Hai and Xu, Peng and Li, Shengjun and Wang, Xiangyu and Guo, Xiangzhou and Li, Chengming and Xu, Xiaohai and Yu, Kun and Yuan, Yuxing and Zou, Yinghao and Long, Jiquan and Cai, Yudong and Li, Zhenxiang and Zhang, Zhifeng and Mo, Yihua and Gu, Jun and Jiang, Ruiyi and Wei, Yi and Xie, Charles},
  booktitle={Proceedings of the 2021 international conference on management of data},
  pages={2614--2627},
  year={2021}
}

@software{jina2025jinarerankerv2,
  title={jina-reranker-v2-base-multilingual},
  author={{Jina AI Team}},
  organization={Jina AI},
  year={2024},
  url={https://huggingface.co/jinaai/jina-reranker-v2-base-multilingual},
  note={Accessed: 2026-04-24}
}

@misc{qwen3technicalreport,
    title={Qwen3 Technical Report}, 
    author={{Qwen Team}},
    year={2025},
    eprint={2505.09388},
    archivePrefix={arXiv},
    primaryClass={cs.CL},
    url={https://arxiv.org/abs/2505.09388}
}

@inproceedings{trivedi2023interleaving,
  title={Interleaving retrieval with chain-of-thought reasoning for knowledge-intensive multi-step questions},
  author={Trivedi, Harsh and Balasubramanian, Niranjan and Khot, Tushar and Sabharwal, Ashish},
  booktitle={Proceedings of the 61st annual meeting of the association for computational linguistics (volume 1: long papers)},
  pages={10014--10037},
  year={2023}
}

@article{yan2024corrective,
  title={Corrective retrieval augmented generation},
  author={Yan, Shi-Qi and Gu, Jia-Chen and Zhu, Yun and Ling, Zhen-Hua},
  journal={arXiv preprint arXiv:2401.15884},
  year={2024}
}

@article{du2026rag,
  title={A-RAG: Scaling Agentic Retrieval-Augmented Generation via Hierarchical Retrieval Interfaces},
  author={Du, Mingxuan and Xu, Benfeng and Zhu, Chiwei and Wang, Shaohan and Wang, Pengyu and Wang, Xiaorui and Mao, Zhendong},
  journal={arXiv preprint arXiv:2602.03442},
  year={2026}
}

@misc{legislation_law_prc_2023,
  title        = {Legislation Law of the People's Republic of China},
  author       = {{National People's Congress of the People's Republic of China}},
  year         = {2023},
  note         = {Adopted on March 15, 2000; amended on March 15, 2015; revised on March 13, 2023},
  howpublished = {\url{https://www.npc.gov.cn/npc/c2/kgfb/202303/t20230314_424438.html}}
}

@inproceedings{nazarenko2018annotation,
  title={An annotation language for semantic search of legal sources},
  author={Nazarenko, Adeline and Levy, Fran{\c{c}}ois and Wyner, Adam},
  booktitle={Proceedings of the Eleventh International Conference on Language Resources and Evaluation (LREC 2018)},
  year={2018}
}

@misc{qwen2.5,
    title = {Qwen2.5: A Party of Foundation Models},
    url = {https://qwenlm.github.io/blog/qwen2.5/},
    author = {{Qwen Team}},
    month = {September},
    year = {2024}
}

@misc{glm2024chatglm,
      title={ChatGLM: A Family of Large Language Models from GLM-130B to GLM-4 All Tools}, 
      author={Team GLM and Aohan Zeng and Bin Xu and Bowen Wang and Chenhui Zhang and Da Yin and Diego Rojas and Guanyu Feng and Hanlin Zhao and Hanyu Lai and Hao Yu and Hongning Wang and Jiadai Sun and Jiajie Zhang and Jiale Cheng and Jiayi Gui and Jie Tang and Jing Zhang and Juanzi Li and Lei Zhao and Lindong Wu and Lucen Zhong and Mingdao Liu and Minlie Huang and Peng Zhang and Qinkai Zheng and Rui Lu and Shuaiqi Duan and Shudan Zhang and Shulin Cao and Shuxun Yang and Weng Lam Tam and Wenyi Zhao and Xiao Liu and Xiao Xia and Xiaohan Zhang and Xiaotao Gu and Xin Lv and Xinghan Liu and Xinyi Liu and Xinyue Yang and Xixuan Song and Xunkai Zhang and Yifan An and Yifan Xu and Yilin Niu and Yuantao Yang and Yueyan Li and Yushi Bai and Yuxiao Dong and Zehan Qi and Zhaoyu Wang and Zhen Yang and Zhengxiao Du and Zhenyu Hou and Zihan Wang},
      year={2024},
      eprint={2406.12793},
      archivePrefix={arXiv},
      primaryClass={cs.CL}
}

@misc{openai2026gpt55,
  title        = {{GPT-5.5}},
  author       = {{OpenAI}},
  year         = {2026},
  howpublished = {\url{https://platform.openai.com/docs/models}},
  note         = {Accessed: 2026-05-23}
}

@misc{anthropic2026claudesonnet46,
  title        = {{Claude Sonnet 4.6}},
  author       = {{Anthropic}},
  year         = {2026},
  howpublished = {\url{https://www.anthropic.com/news/claude-sonnet-4-6}},
  note         = {Accessed: 2026-05-23}
}

@inproceedings{su2024stard,
  title={STARD: A Chinese Statute Retrieval Dataset Derived from Real-life Queries by Non-professionals},
  author={Su, Weihang and Hu, Yiran and Xie, Anzhe and Ai, Qingyao and Bing, Quezi and Zheng, Ning and Liu, Yun and Shen, Weixing and Liu, Yiqun},
  booktitle={Findings of the Association for Computational Linguistics: EMNLP 2024},
  pages={10658--10671},
  year={2024}
}

@article{pipitone2024legalbench,
  title={Legalbench-rag: A benchmark for retrieval-augmented generation in the legal domain},
  author={Pipitone, Nicholas and Alami, Ghita Houir},
  journal={arXiv preprint arXiv:2408.10343},
  year={2024}
}

@article{yu2022legal,
  title={Legal prompting: Teaching a language model to think like a lawyer},
  author={Yu, Fangyi and Quartey, Lee and Schilder, Frank},
  journal={arXiv preprint arXiv:2212.01326},
  year={2022}
}

@inproceedings{servantez2024chain,
  title = {Chain of Logic: Rule-Based Reasoning with Large Language Models},
  author = {Servantez, Sergio and Barrow, Joe and Hammond, Kristian and Jain, Rajiv},
  booktitle = {Findings of the Association for Computational Linguistics: ACL 2024},
  year = {2024},
  pages = {2721--2733},
  publisher = {Association for Computational Linguistics}
}

@inproceedings{li2025delta,
  title={Delta: Pre-train a discriminative encoder for legal case retrieval via structural word alignment},
  author={Li, Haitao and Ai, Qingyao and Han, Xinyan and Chen, Jia and Dong, Qian and Liu, Yiqun},
  booktitle={Proceedings of the AAAI Conference on Artificial Intelligence},
  volume={39},
  number={25},
  pages={27072--27080},
  year={2025}
}

@article{shao2023intent,
  title={An intent taxonomy of legal case retrieval},
  author={Shao, Yunqiu and Li, Haitao and Wu, Yueyue and Liu, Yiqun and Ai, Qingyao and Mao, Jiaxin and Ma, Yixiao and Ma, Shaoping},
  journal={ACM Transactions on Information Systems},
  volume={42},
  number={2},
  pages={1--27},
  year={2023},
  publisher={ACM New York, NY, USA}
}

@inproceedings{ponte2017language,
  title={A language modeling approach to information retrieval},
  author={Ponte, Jay M and Croft, W Bruce},
  booktitle={ACM SIGIR Forum},
  volume={51},
  number={2},
  pages={202--208},
  year={2017},
  organization={ACM New York, NY, USA}
}

@article{robertson2009probabilistic,
  title={The probabilistic relevance framework: BM25 and beyond},
  author={Robertson, Stephen and Zaragoza, Hugo},
  journal={Foundations and Trends in Information Retrieval},
  volume={3},
  number={4},
  pages={333--389},
  year={2009},
  doi={10.1561/1500000019}
}

@misc{text2vec,
  author = {Xu, Ming},
  title = {Text2vec: Text to vector toolkit},
  year = {2022},
  publisher = {GitHub},
  journal = {GitHub repository},
  howpublished = {\url{https://github.com/shibing624/text2vec}}
}

@misc{bge_embedding,
      title={C-Pack: Packaged Resources To Advance General Chinese Embedding}, 
      author={Shitao Xiao and Zheng Liu and Peitian Zhang and Niklas Muennighoff},
      year={2023},
      eprint={2309.07597},
      archivePrefix={arXiv},
      primaryClass={cs.CL}
}

@misc{bge-m3,
      title={BGE M3-Embedding: Multi-Lingual, Multi-Functionality, Multi-Granularity Text Embeddings Through Self-Knowledge Distillation}, 
      author={Jianlv Chen and Shitao Xiao and Peitian Zhang and Kun Luo and Defu Lian and Zheng Liu},
      year={2024},
      eprint={2402.03216},
      archivePrefix={arXiv},
      primaryClass={cs.CL}
}

@article{li2023towards,
  title={Towards general text embeddings with multi-stage contrastive learning},
  author={Li, Zehan and Zhang, Xin and Zhang, Yanzhao and Long, Dingkun and Xie, Pengjun and Zhang, Meishan},
  journal={arXiv preprint arXiv:2308.03281},
  year={2023}
}

@article{wang2024multilingual,
  title={Multilingual E5 Text Embeddings: A Technical Report},
  author={Wang, Liang and Yang, Nan and Huang, Xiaolong and Yang, Linjun and Majumder, Rangan and Wei, Furu},
  journal={arXiv preprint arXiv:2402.05672},
  year={2024}
}

@misc{ChatLaw,
  author={Jiaxi Cui and Zongjian Li and Yang Yan and Bohua Chen and Li Yuan},
  title={ChatLaw},
  year={2023},
  publisher={GitHub},
  journal={GitHub repository},
  howpublished={\url{https://github.com/PKU-YuanGroup/ChatLaw}},
}

@inproceedings{SAILER,
      title={SAILER: Structure-aware Pre-trained Language Model for Legal Case Retrieval}, 
      author={Haitao Li and Qingyao Ai and Jia Chen and Qian Dong and Yueyue Wu and Yiqun Liu and Chong Chen and Qi Tian},
      booktitle={Proceedings of the 46th International ACM SIGIR Conference on Research and Development in Information Retrieval},
      pages={1035--1044},
      year={2023}
}

@inproceedings{li2025lexrag,
  title={LexRAG: Benchmarking retrieval-augmented generation in multi-turn legal consultation conversation},
  author={Li, Haitao and Chen, Yifan and YiRan, Hu and Ai, Qingyao and Chen, Junjie and Yang, Xiaoyu and Yang, Jianhui and Wu, Yueyue and Liu, Zeyang and Liu, Yiqun},
  booktitle={Proceedings of the 48th International ACM SIGIR Conference on Research and Development in Information Retrieval},
  pages={3606--3615},
  year={2025}
}

@book{shanghai_market_q&a_2021,
  author    = {{Shanghai Municipal Administration for Market Regulation}},
  title     = {Market Supervision and Administration Practical Training:
               Analysis of 500 Questions},
  publisher = {China Industry and Commerce Press},
  address   = {Beijing},
  year      = {2021},
  isbn      = {978-7-5209-0148-2}
}

@article{xiao2018cail2018,
  title={Cail2018: A large-scale legal dataset for judgment prediction},
  author={Xiao, Chaojun and Zhong, Haoxi and Guo, Zhipeng and Tu, Cunchao and Liu, Zhiyuan and Sun, Maosong and Feng, Yansong and Han, Xianpei and Hu, Zhen and Wang, Heng and Xu, Jianfeng},
  journal={arXiv preprint arXiv:1807.02478},
  year={2018}
}

@inproceedings{fei2024lawbench,
  title={Lawbench: Benchmarking legal knowledge of large language models},
  author={Fei, Zhiwei and Shen, Xiaoyu and Zhu, Dawei and Zhou, Fengzhe and Han, Zhuo and Huang, Alan and Zhang, Songyang and Chen, Kai and Yin, Zhixin and Shen, Zongwen and Ge, Jidong and Ng, Vincent},
  booktitle={Proceedings of the 2024 conference on empirical methods in natural language processing},
  pages={7933--7962},
  year={2024}
}

@inproceedings{li2025legalagentbench,
  title={Legalagentbench: Evaluating llm agents in legal domain},
  author={Li, Haitao and Chen, Junjie and Yang, Jingli and Ai, Qingyao and Jia, Wei and Liu, Youfeng and Lin, Kai and Wu, Yueyue and Yuan, Guozhi and Hu, Yiran and Wang, Wuyue and Liu, Yiqun and Huang, Minlie},
  booktitle={Proceedings of the 63rd Annual Meeting of the Association for Computational Linguistics (Volume 1: Long Papers)},
  pages={2322--2344},
  year={2025}
}

@inproceedings{su2025judge,
  title={Judge: Benchmarking judgment document generation for chinese legal system},
  author={Su, Weihang and Yue, Baoqing and Ai, Qingyao and Hu, Yiran and Li, Jiaqi and Wang, Changyue and Zhang, Kaiyuan and Wu, Yueyue and Liu, Yiqun},
  booktitle={Proceedings of the 48th International ACM SIGIR Conference on Research and Development in Information Retrieval},
  pages={3573--3583},
  year={2025}
}

@inproceedings{bhattacharya2019fire,
  title={FIRE 2019 AILA track: artificial intelligence for legal assistance},
  author={Bhattacharya, Paheli and Ghosh, Kripabandhu and Ghosh, Saptarshi and Pal, Arindam and Mehta, Parth and Bhattacharya, Arnab and Majumder, Prasenjit},
  booktitle={Proceedings of the 11th annual meeting of the forum for information retrieval evaluation},
  pages={4--6},
  year={2019}
}

@inproceedings{louis2022statutory,
  title={A statutory article retrieval dataset in French},
  author={Louis, Antoine and Spanakis, Gerasimos},
  booktitle={Proceedings of the 60th Annual Meeting of the Association for Computational Linguistics (Volume 1: Long Papers)},
  pages={6789--6803},
  year={2022}
}

@inproceedings{goebel2023summary,
  title={Summary of the competition on legal information, extraction/entailment (COLIEE) 2023},
  author={Goebel, Randy and Kano, Yoshinobu and Kim, Mi-Young and Rabelo, Juliano and Satoh, Ken and Yoshioka, Masaharu},
  booktitle={Proceedings of the nineteenth international conference on artificial intelligence and law},
  pages={472--480},
  year={2023}
}

@inproceedings{hou2025clerc,
  title={CLERC: A dataset for US legal case retrieval and retrieval-augmented analysis generation},
  author={Hou, Abe Bohan and Weller, Orion and Qin, Guanghui and Yang, Eugene and Lawrie, Dawn and Holzenberger, Nils and Blair-Stanek, Andrew and Van Durme, Benjamin},
  booktitle={Findings of the Association for Computational Linguistics: NAACL 2025},
  pages={7913--7928},
  year={2025}
}

@inproceedings{kim2025legalsearchlm,
  title={Legalsearchlm: Rethinking legal case retrieval as legal elements generation},
  author={Kim, Chaeeun and Lee, Jinu and Hwang, Wonseok},
  booktitle={Proceedings of the 2025 Conference on Empirical Methods in Natural Language Processing},
  pages={4521--4554},
  year={2025}
}

@inproceedings{deng2024element,
  title={An element is worth a thousand words: Enhancing legal case retrieval by incorporating legal elements},
  author={Deng, Chenlong and Dou, Zhicheng and Zhou, Yujia and Zhang, Peitian and Mao, Kelong},
  booktitle={Findings of the Association for Computational Linguistics: ACL 2024},
  pages={2354--2365},
  year={2024}
}

@inproceedings{ma2023query,
  title={Query rewriting in retrieval-augmented large language models},
  author={Ma, Xinbei and Gong, Yeyun and He, Pengcheng and Zhao, Hai and Duan, Nan},
  booktitle={Proceedings of the 2023 Conference on Empirical Methods in Natural Language Processing},
  pages={5303--5315},
  year={2023}
}

@inproceedings{wang2023query2doc,
  title={Query2doc: Query expansion with large language models},
  author={Wang, Liang and Yang, Nan and Wei, Furu},
  booktitle={Proceedings of the 2023 Conference on Empirical Methods in Natural Language Processing},
  pages={9414--9423},
  year={2023}
}

@inproceedings{gao2023precise,
  title={Precise zero-shot dense retrieval without relevance labels},
  author={Gao, Luyu and Ma, Xueguang and Lin, Jimmy and Callan, Jamie},
  booktitle={Proceedings of the 61st Annual Meeting of the Association for Computational Linguistics (Volume 1: Long Papers)},
  pages={1762--1777},
  year={2023}
}

@inproceedings{kwon2023efficient,
  title = {Efficient Memory Management for Large Language Model Serving with PagedAttention},
  author = {Kwon, Woosuk and Li, Zhuohan and Zhuang, Siyuan and Sheng, Ying and Zheng, Lianmin and Yu, Cody Hao and Gonzalez, Joseph E. and Zhang, Hao and Stoica, Ion},
  booktitle = {Proceedings of the 29th Symposium on Operating Systems Principles},
  pages = {611--626},
  year = {2023}
}

@misc{openai_python,
  title = {OpenAI Python API Library},
  author = {{OpenAI}},
  year = {2026},
  howpublished = {\url{https://github.com/openai/openai-python}},
  note = {Accessed: 2026-05-13}
}

\appendix

\section{Prompt for \textsc{LexPath}}

\begin{figure}
    \centering    \includegraphics[width=0.95\linewidth,trim = 0 400 0 0]{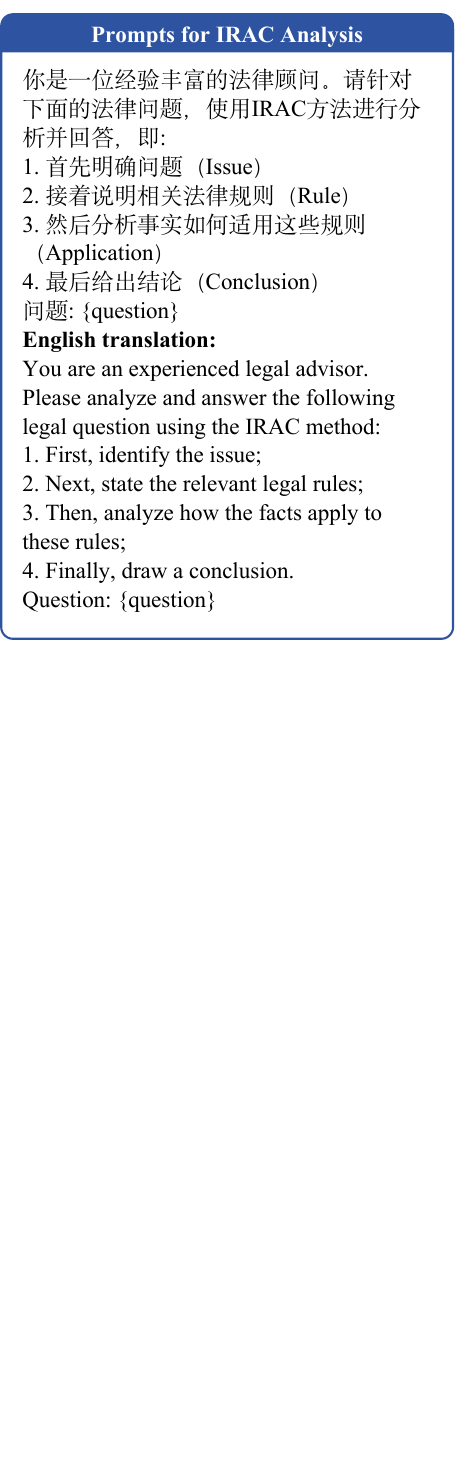}
    \caption{Prompt for IRAC analysis.}
    \label{fig:prompt-irac}
\end{figure}

\begin{figure}
    \centering    \includegraphics[width=0.95\linewidth,trim = 0 170 0 0]{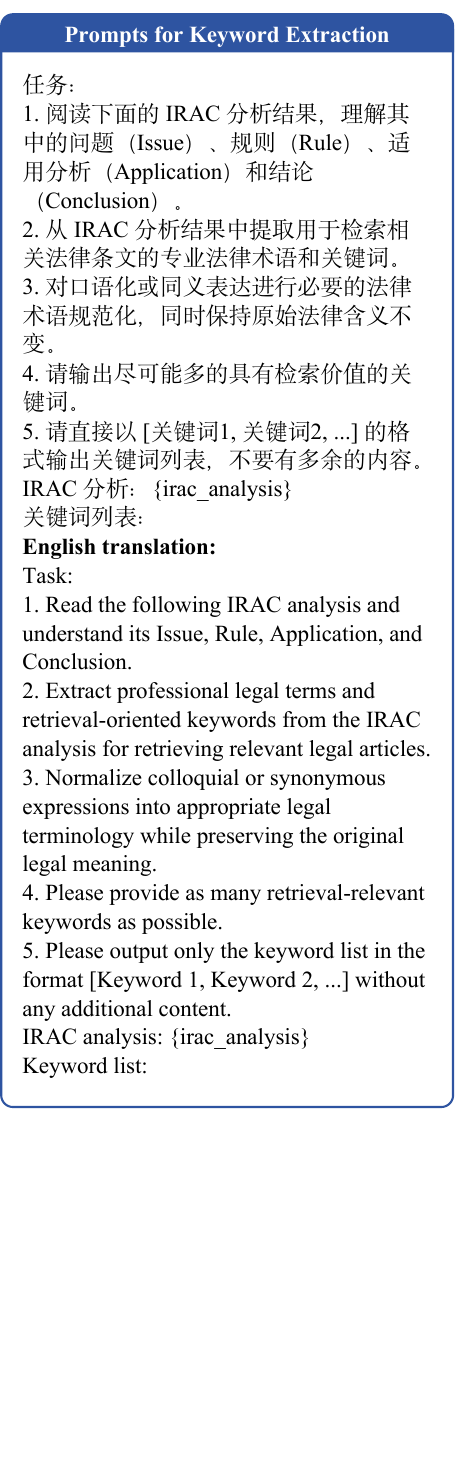}
    \caption{Prompt for keyword extraction.}
    \label{fig:prompt-keyword}
\end{figure}

\begin{figure*}
    \centering    \includegraphics[width=0.95\linewidth,trim = 0 110 0 0]{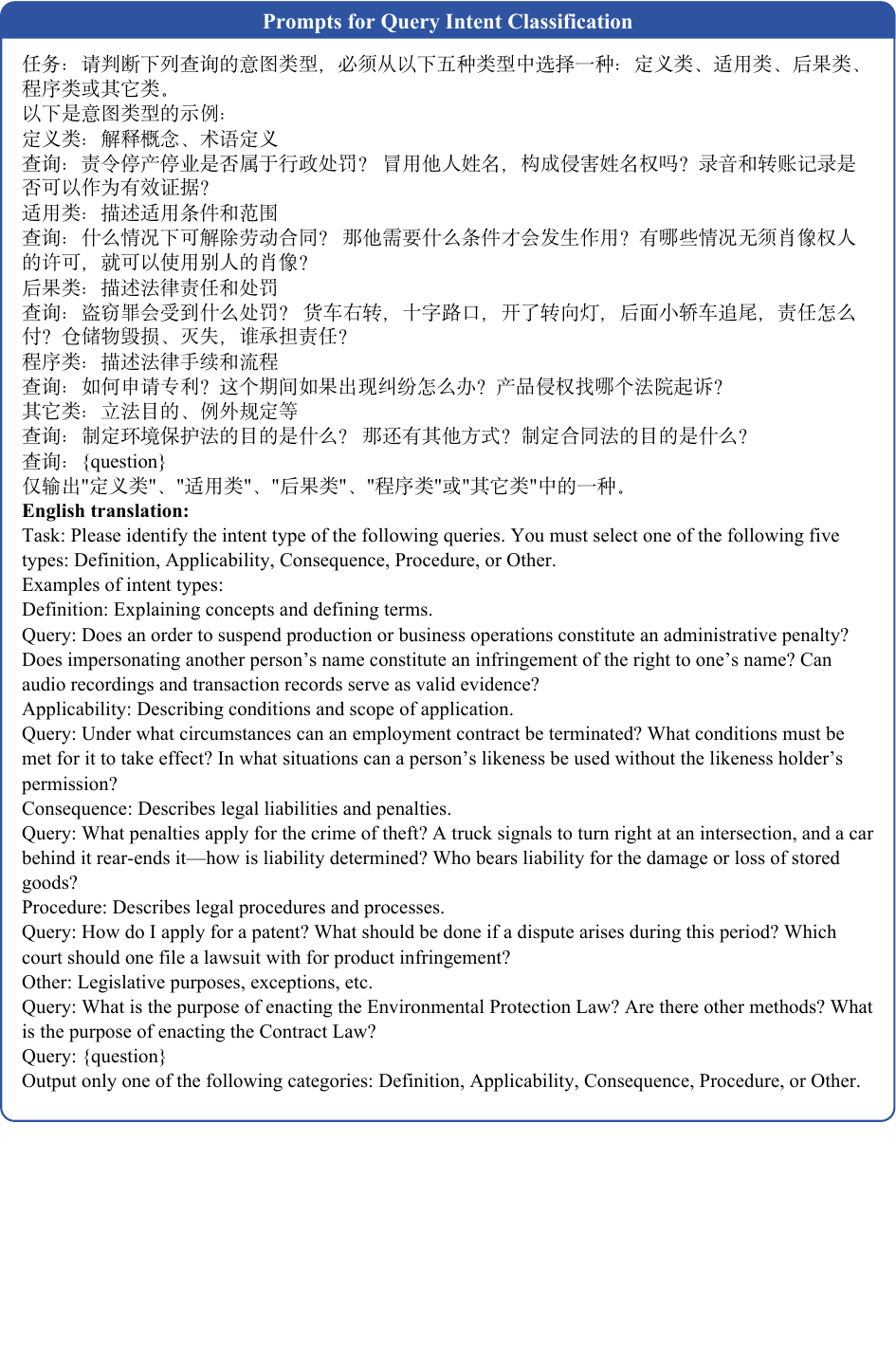}
    \caption{Prompt for query intent classification.}
    \label{fig:prompt-query}
\end{figure*}

\begin{figure*}
    \centering    \includegraphics[width=0.95\linewidth,trim = 0 10 0 100]{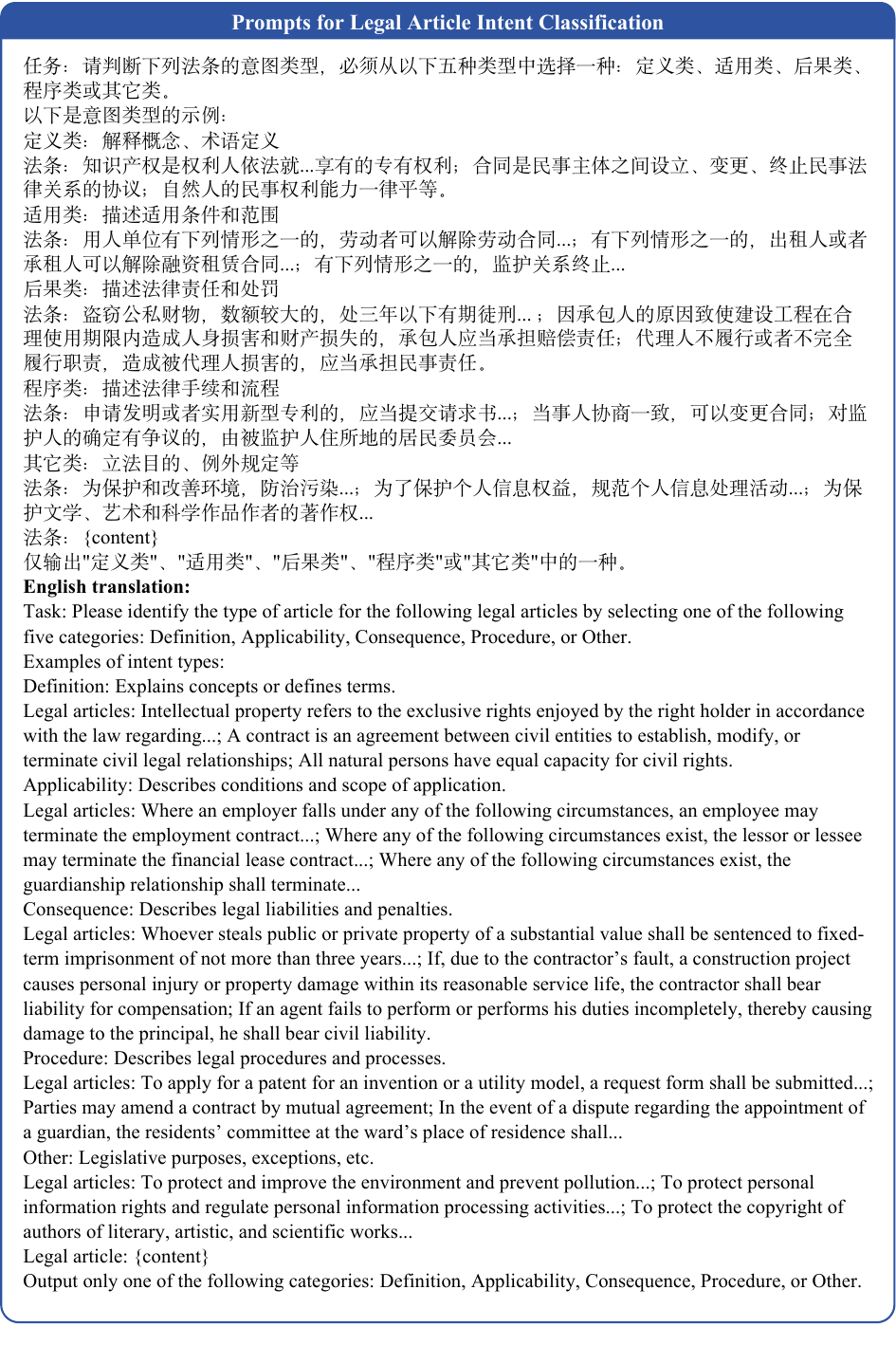}
    \caption{Prompt for legal article intent classification.}
    \label{fig:prompt-article}
\end{figure*}

The prompts used in IRAC-Exp are shown in Figure~\ref{fig:prompt-irac} for IRAC analysis and Figure~\ref{fig:prompt-keyword} for keyword extraction.
The prompts used for few-shot intent classification are shown in Figure~\ref{fig:prompt-query} for queries and Figure~\ref{fig:prompt-article} for legal articles.

\label{sec:prompt}

\section{Construction Process of StatuteRAG}
\label{sec:statuterag}
Existing benchmarks of legal article retrieval~\cite{su2024stard,li2025lexrag} are collected from daily legal consultations submitted by the general public.
To supplement the application scenarios for the legal article retrieval task, we propose StatuteRAG, in which queries are adapted from questions used to train domain professionals.

The seed questions to build StatuteRAG are selected from a publicly available manual~\cite{shanghai_market_q&a_2021} used for training legal professionals.
The manual focuses on market supervision and administration laws and contains 500 single-choice questions, together with explanatory analyses and legal article references for each option.
The process of constructing a query dataset is as follows:
\begin{itemize}
    \item Exclude questions that are overly trivial, ambiguous, overly context-dependent, have incomplete context, or lack a clear legal article reference, and use the remaining ones as seed questions.
    \item For each seed question, use regular expressions to extract the text from the question stem and each answer choice.
    \item Combine the question stem with the content of each answer choice to create a true-or-false question, and match it with the corresponding legal articles.
\end{itemize}

To ensure data quality, the constructed samples are manually reviewed.
Since StatuteRAG is adapted from a professional training manual with existing explanations and legal article references, the review focuses on verifying the conversion quality, including query completeness, answer consistency, and article support.
Samples with unclear wording, missing context, or weak article support are removed.

The resulting dataset contains 1,361 queries annotated with relevant legal articles and ground-truth answers, together with a corpus of 56,982 candidate legal articles compiled from 1,295 effective legal documents at the time of collection.
The statistics of StatuteRAG are shown in Table~\ref{tab:statuterag}.

\begin{table}[h]
\centering
\small
\caption{Statistics of StatuteRAG.}
\begin{tabular}{lc}
\toprule
Statistic & Value \\
\midrule
Number of queries & 1,361 \\
Number of candidate legal articles & 56,982 \\
Number of referenced article annotations & 1,406 \\
Average number of articles referenced per query & 1.03 \\
Average query length & 72.57 \\
Average legal article length & 86.90 \\
\bottomrule
\end{tabular}
\label{tab:statuterag}
\end{table}

\section{Challenges on LexRAG}
\label{sec:chall}
LexRAG is challenging because its queries are derived from real multi-turn legal consultations.
Compared with STARD and StatuteRAG, the information needed for retrieval is not always contained in a single self-contained query.
Relevant facts, legal issues, and user intents may be distributed across multiple dialogue turns, which partly explains the relatively lower recall on LexRAG.

First, LexRAG requires retrievers to handle dialogue-dependent information.
Key facts may appear in earlier turns rather than in the final user question alone.
In the following example, the last turn asks how the user can protect their rights, but the underlying issue depends on previous turns about a financed car, insurance requirements, and vehicle repossession.

{\footnotesize\itshape
Turn 1: If I did not purchase the second-year insurance for my financed car through the finance company, do they have the right to repossess the vehicle? I have been making my monthly payments on time.

Turn 2: Should we specify the method of purchasing insurance in this agreement?

Turn 3: Is it legal for a finance company to repossess a vehicle?

Turn 4: Is there any other way for this company to hold the other party accountable?

Turn 5: What can I do to protect my rights if he forcibly impounds my vehicle?
}

Second, LexRAG contains many queries from ordinary users, which increases the gap between user expressions and legal language.
Users often describe concrete facts in everyday language without explicitly mentioning the legal concepts useful for retrieval.
For example, the following query describes a lost phone and surveillance footage, but does not directly state concepts such as evidence collection, privacy protection, or access to records.

{\footnotesize\itshape
I lost my phone in the classroom this morning, but when I went to the school to check the surveillance footage, they refused to let me.
}

Third, the legal intent of a LexRAG query can be implicit or underspecified.
In the following example, the user asks whether parents can give resettlement housing to their grandson without a property deed, while the reference article concerns the effectiveness of real property rights through registration.
Although the query and the article are legally related, the connection is indirect and may not be captured by a single intent label.

{\footnotesize\itshape
Query: Hello, my parents would like to give their resettlement housing to their grandson, but they currently do not have a property deed.

Reference Article: Article 209 of the Civil Code of the People's Republic of China: The creation, modification, transfer, and termination of real property rights take effect upon registration in accordance with the law; they do not take effect without such registration, unless otherwise provided by law. Ownership of natural resources that belong to the State by law may be exempt from registration.
}

These examples suggest that LexRAG requires more context-aware retrieval and more flexible intent modeling.
We leave the integration of dialogue context and multi-intent retrieval as future work.

\section{Impact on Downstream Legal QA}
\label{sec:legalqa}
To assess the impact of retrieval quality on downstream legal QA, we evaluate two lightweight models, GLM4-9B~\cite{glm2024chatglm} and Qwen2.5-7B-Instruct~\cite{qwen2.5}, together with two flagship models, Claude4.6-Sonnet~\cite{anthropic2026claudesonnet46} and GPT-5.5~\cite{openai2026gpt55}, under four retriever settings: zero-shot, BM25, bge-large-zh-v1.5, and \textsc{LexPath}. Following~\cite{li2025lexrag}, keyword accuracy is adopted as the metric for short-answer queries in LexRAG.
For StatuteRAG, which consists entirely of true-or-false queries, answer accuracy is reported. STARD is excluded because it does not provide reference answers.

Experimental results in Table~\ref{tab:legalqa} suggest the following findings:
(1) Model scale alone does not guarantee stronger performance in domain-specific QA, as flagship models do not consistently outperform lightweight models in the zero-shot setting.
(2) \textsc{LexPath} generally brings larger gains than BM25 and bge-large-zh-v1.5, showing the value of domain-oriented retrieval design.
(3) The improvements are especially clear for lightweight models and StatuteRAG, indicating that high-quality retrieval can benefit resource-constrained legal QA scenarios.
(4) For flagship models on LexRAG, retrieval-augmented results are less stable, possibly because limited retrieval recall introduces noisy or incomplete evidence. 

\begin{table}[h]
\centering
\small
\setlength{\tabcolsep}{4pt}
\renewcommand{\arraystretch}{0.95}
\caption{Performance of different retriever settings on downstream legal QA. ``BGE'' denotes bge-large-zh-v1.5.}
\begin{tabular}{l@{\hspace{5pt}}|@{\hspace{5pt}}ccc}
\toprule
\textbf{Generator} & \textbf{Retriever} & \textbf{LexRAG} & \textbf{StatuteRAG}\\
\midrule
\multirow{4}{*}{GLM4-9B} 
& Zero-Shot & 22.49 & 62.64 \\
& BM25 & 23.77 & \underline{77.29} \\
& BGE & \underline{28.21} & 75.09 \\
& \textsc{LexPath} & \textbf{30.18} & \textbf{78.39} \\
\midrule
\multirow{4}{*}{Qwen2.5-7B} 
& Zero-Shot & \underline{31.07} & 76.19 \\
& BM25 & 27.81 & \underline{76.56}\\
& BGE & 27.51 & 75.46\\
& \textsc{LexPath} & \textbf{33.53} & \textbf{78.75} \\
\midrule
\multirow{4}{*}{Claude4.6-Sonnet} 
& Zero-Shot & \textbf{33.33} & 67.77 \\
& BM25 & 25.25 & 76.92\\
& BGE & 27.32 & 72.16 \\
& \textsc{LexPath} & \underline{29.09} & \textbf{86.81} \\
\midrule
\multirow{4}{*}{GPT-5.5} 
& Zero-Shot & \textbf{32.25} & 80.59 \\
& BM25 & 23.77 & \underline{84.25}\\
& BGE & 27.51 & 82.78 \\
& \textsc{LexPath} & \underline{28.01} & \textbf{87.91} \\
\bottomrule
\end{tabular}
\label{tab:legalqa}
\end{table}

\section{LLM Error Analysis}
\label{sec:llm-error-analysis}

\textsc{LexPath} uses three online LLM calls: IRAC analysis,
keyword extraction from the IRAC output, and query intent prediction.
We further analyze how errors from these components may affect retrieval
and reranking.

\paragraph{Errors in IRAC-based Query Expansion}
The LLM may generate incorrect, outdated, or nonexistent article
references during IRAC analysis. Such errors may be propagated to the
subsequently extracted keywords and introduce noise into sparse retrieval.
For example, for the query ``In a financial loan, how should the time for
interest payment be determined?'', IRAC-Exp generates useful expressions
such as \textit{term}, \textit{agreement}, \textit{repayment of the loan},
and \textit{at the end of each full year}, which occur in the ground-truth
Article 674 of the Civil Code. However, the generated output also contains
additional article references that do not correspond to the
ground-truth article such as ``Contract Law, Article 205'' and
``Article 61''. Despite such noise, the generated keywords are appended
to the original query rather than replacing it, and the dense retrieval
path is independent of IRAC-Exp. Therefore, errors in query expansion do
not directly determine the final retrieval results.

\paragraph{Errors in Intent Prediction}
Incorrect intent predictions may change the intent-consistency score and
thereby affect the ordering of candidate articles. However, intent
consistency is only one component of the final reranking score, together
with the original retrieval score and the neural reranker score over the
top-20 candidates. Consequently, an incorrect intent label may influence
the ranking but does not determine it alone.

\paragraph{Reranking Failure Analysis}
We further examine cases in which reranking degrades retrieval quality.
Among queries whose gold article appears in the top 5 before reranking,
only 1.62\% on STARD, 2.27\% on LexRAG, and 1.10\% on StatuteRAG have
their gold article moved out of the top 5 after reranking.
We identify three main causes. First, the query and the gold article may
have different dominant intents, causing the intent-consistency signal to
favor another candidate. Second, an incorrect intent prediction may alter
the consistency score. Third, the general-purpose neural reranker may miss
fine-grained legal distinctions, especially for short queries.

\begin{table}[t]
\centering
\small
\setlength{\tabcolsep}{3pt}
\caption{Representative cases where reranking lowers the rank of the
gold article.}
\label{tab:reranking-errors}
\begin{tabularx}{\columnwidth}{@{}p{0.3\columnwidth}Xp{0.16\columnwidth}@{}}
\toprule
\textbf{Error Type} &
\textbf{Query} &
\textbf{Rank Change} \\
\midrule

Intent mismatch
& How will gambling debts be handled in a divorce?
& $4 \rightarrow 8$ \\

Intent prediction
& What does the law stipulate regarding the process for disability assessment?
& $4 \rightarrow 6$ \\

Neural reranker
& What are public funds?
& $4 \rightarrow 7$ \\

\bottomrule
\end{tabularx}
\end{table}

Overall, LLM-related errors mainly affect the sparse retrieval path and
intent-aware reranking. The independent dense path and the combination
of multiple ranking signals reduce the dependence of the final result on
any single LLM prediction.

\section{Case Study}
\label{sec:case}
\begin{figure*}
    \centering
    \includegraphics[width=0.9\linewidth,trim = 150 0 150 0]{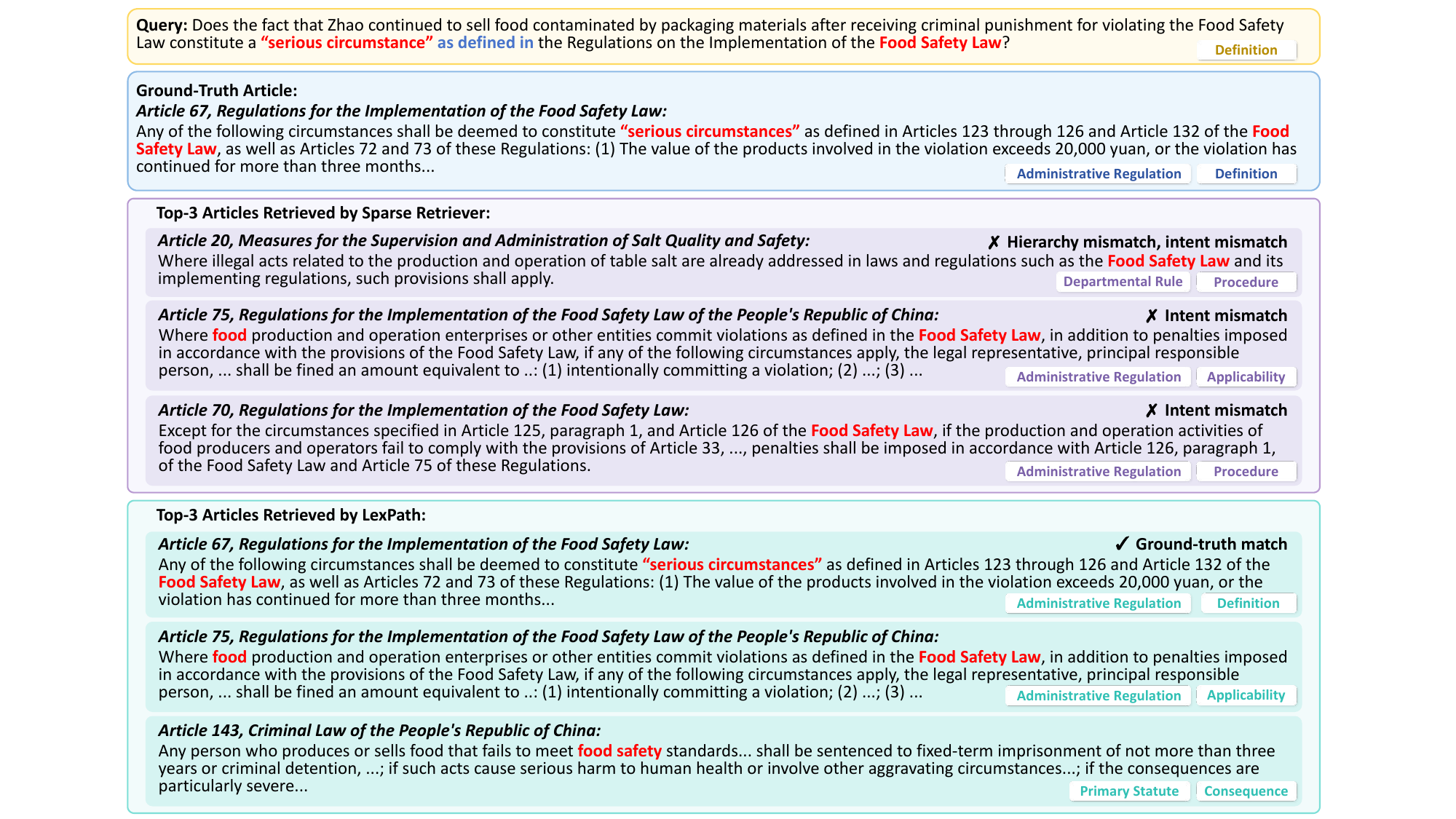}
    \caption{Case study of a standalone sparse retrieval failure. The sparse retriever ranks articles with keyword overlap but mismatched hierarchy or intent, while \textsc{LexPath} retrieves the ground-truth definitional article at the top.}
    \label{fig:case1}
\end{figure*}

Figure~\ref{fig:case1} shows a typical failure case of sparse retrieval. 
The query asks whether Zhao's conduct constitutes a ``serious circumstance'' under the Regulations for the Implementation of the Food Safety Law, so the ground-truth article is a definitional article that directly explains this term. 
However, the sparse retriever ranks articles that contain overlapping keywords such as ``Food Safety Law'' and ``food'', but these articles either come from a mismatched legal hierarchy or focus on applicability and procedure rather than definition. 
In contrast, \textsc{LexPath} ranks the ground-truth article first. 
This case shows that keyword overlap alone is not enough for legal article retrieval, and that hierarchy and intent information help identify the article that directly supports the query.

\begin{figure*}
    \centering
    \includegraphics[width=0.9\linewidth,trim = 150 0 150 0]{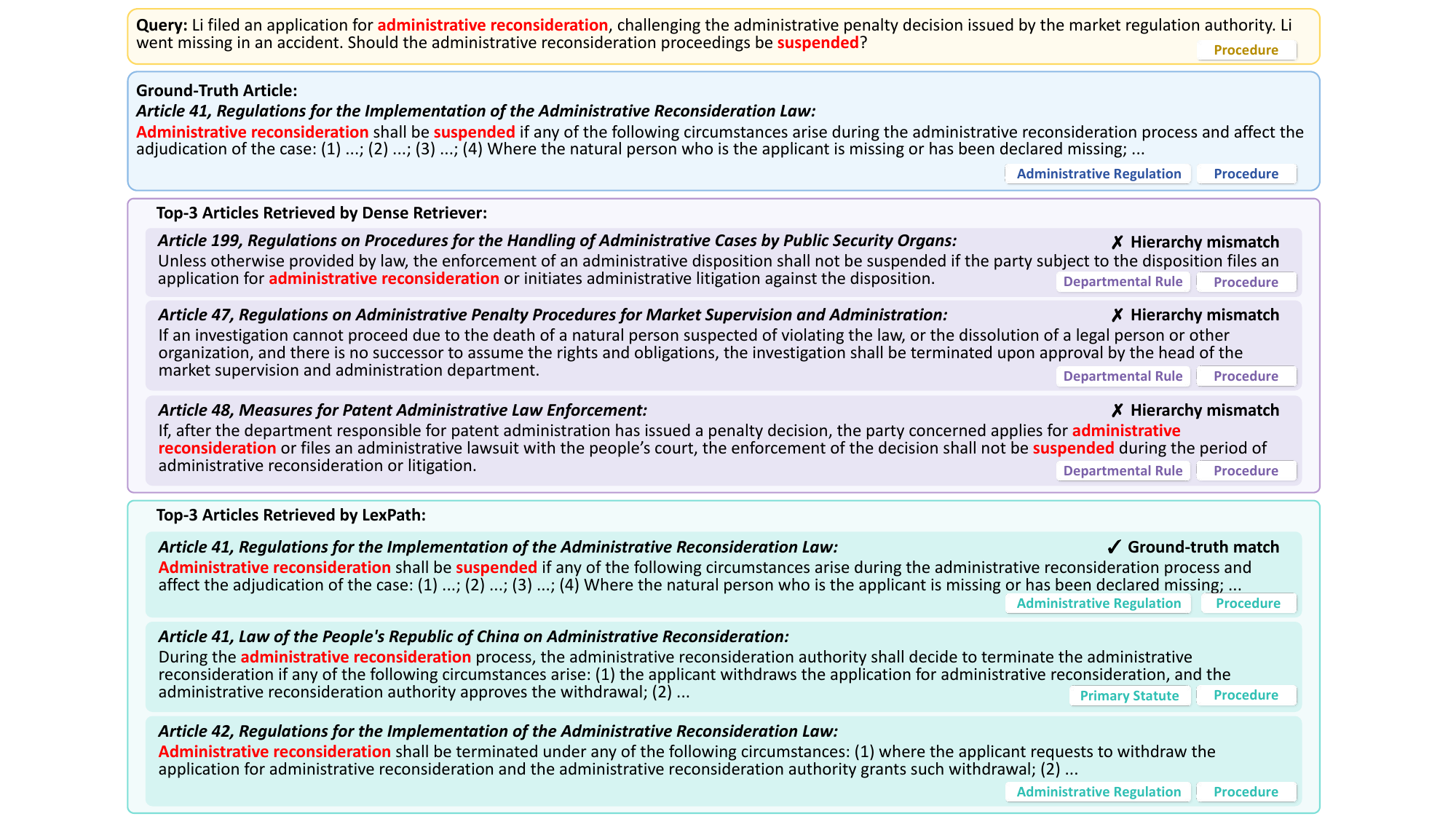}
    \caption{Case study of a standalone dense retrieval failure. The dense retriever retrieves semantically related procedural articles but fails to distinguish legal hierarchy and the key condition of suspension, while \textsc{LexPath} ranks the ground-truth article first.}
    \label{fig:case2}
\end{figure*}

Figure~\ref{fig:case2} shows a typical dense retrieval error. 
The query asks whether the reconsideration process should be suspended after the applicant goes missing. 
The dense retriever returns articles that are related to administrative reconsideration or administrative procedures, but these articles mainly concern other situations, such as non-suspension of enforcement or termination of investigation. 
They also come from lower-level departmental rules. 
\textsc{LexPath} ranks the ground-truth article first, which directly states that reconsideration shall be suspended when the applicant is missing. 
This case shows that legal retrieval needs to distinguish not only semantic similarity, but also legal hierarchy and the exact procedural condition.

\end{document}